# Service Cloaking and Authentication at Data Link Layer

Arun Kumar S P

Centre for Artificial Intelligence and Robotics

Bangalore

arunkumarsp@fastmail.fm

**Abstract--This paper discusses that there is significant benefit in providing stronger security at lower layers of the network stack for hosts connected to a network. It claims to reduce the attack vulnerability of a networked host by providing security mechanisms in a programmable Network Interface Card (NIC). Dynamic access control mechanisms are implemented in hardware to restrict access to the services provided, only to authenticated hosts. This reduces server vulnerability to various layer 2 attacks. Also the services will be immune to zero-day vulnerabilities due to the minimal code execution paths. To this end, it presents architecture and implementation details of a programmable network interface card equipped with these measures. It works alongside, and augments, existing security protocols making deployment practical.**

## I INTRODUCTION

Network security has become a concern with the rapid growth of the Internet. It plays one of the key roles in the network infrastructure of an organization. Although a lot of research has been carried out to provide security in application, transport and network layers [1, 2, 3], the importance of security in data link layer is still underestimated. Traditional approaches to network security are not found to be effective in local area networks (LANs). A data link layer communication is a weak link in terms of network security. Since a system is only as strong as it's weakest link, a compromise at this level will invalidate the other levels of security. A Denial of Service (DoS) [4] attack using ARP cache poisoning [5] can render a server useless even if it provides other forms of security. Thus, the security measures at the data link layer are complementary to the other measures that are used to provide extra protection of the network and users.

Another appealing factor about data link layer security is that it can conceal the vulnerabilities present in the upper layers to the outside world. One of the major difficulties in protecting networked machines is because of the vulnerabilities present in the operating system and services running in the host. A number of exploit releases, worms and vulnerability disclosures [6, 7, 8, 9] allow an attacker to exploit them to get access into the system. Although secure code writing practices are followed these days, the vicious 'exploit-disclosure-patch' cycle seems to be unavoidable. Hardening the system [10] by restricting the number of open ports and avoiding information leakage are only temporary solutions, since the open ports too can have vulnerabilities. Moreover the operating system which includes the network stack is not devoid of vulnerabilities.

This paper considers the security challenges in IP over Ethernet networks because of their predominance. Ethernet provides a low-cost, high-speed, general purpose communication interface. It is now standardized by IEEE [11] and is available in various data rates. Although Ethernet is considered the concepts proposed are general.

In this paper we propose a network security model that eliminates the various data link layer security issues and also reduces the attack vulnerability of a network server. We propose to apply necessary security measures in the network interface devices. We describe the implementation details using a Programmable Network Interface Card based on Field Programmable Gate Array (FPGA).

This paper is organized as follows. We begin with the problems considered in Section 2. The architecture and its

FPGA based implementation are discussed in Section 3 and 4, respectively. Related work is described in Section 5 Summary of the contributions and conclusions are presented in section 6.

## II NETWORK SECURITY

Weakness in data link layer coupled with programming flaws in the networking software form the fundamental problems in network security.

*A. Data Link Layer Attacks*

An Ethernet network has many security weaknesses that can be used to launch attacks from inside or outside the network. Attacks that are targeted towards end systems exploit the disadvantages of Ethernet which are discussed in [12]. The primary weakness with Ethernet is that it is a broadcast system. Every message reaches all parts of the LAN and can be read by any host on the network. This allows an attacker to passively eavesdrop packets going through the network. Also, Ethernet does not provide any mechanism to authenticate the senders' identity. This allows an attacker to generate spoofed packets or to replay earlier eavesdropped packets.

Most commonly known attacks against end hosts IP over Ethernet networks are based on Medium Access Control (MAC) Address spoofing or Address Resolution Protocol (ARP) poisoning. MAC spoofing attacks involve the use of a known MAC address of another host to attempt to make the target switch forward frames destined for the remote host to the network attacker. By sending a single frame with the other host's source Ethernet address, the network attacker overwrites the table entry so that the switch forwards packets destined for the host to the network attacker. Until the host sends traffic it will not receive any traffic. When the host sends out traffic, the table entry is rewritten once more so that it moves back to the original port.

ARP Poisoning attack exploits the weakness in the design of ARP. ARP [13] is a protocol that provides the mapping between the 32 bit IP address and the 48 bit Ethernet address. An ARP request packet is broadcasted whenever a host needs to send an IP datagram as an Ethernet frame for the first time. Host with the matching IP address replies by a unicast packet to the sender of the request with the <IP, MAC> pairs. To minimize the number of requests sent on the network every host maintains a table of <IP, MAC> pairs, called the ARP cache. This table is filled based on the replies received and the entries have an expiration time. Since ARP is a stateless protocol; a reply is processed even though the corresponding request was never transmitted. This allows an attacker to easily change the <IP, MAC> association contained in a host ARP cache by forging an ARP reply packet. This forms the basis of many attacks like the DoS and Man-In-The-Middle (MITM) [5].

*B. Software Vulnerabilities*

The security of a system depends on a wide variety of configuration elements both at the operating system level and the application level. As the old saying goes, "The devil is in the details." Software services and the operating system are nearly infinitely configurable, and subtle configuration changes can have significant security implications. Thus, some security exposures and vulnerabilities are not always immediately obvious, and a lack of understanding about the impact of changing configuration elements can lead to unintentional exposures.

Furthermore, security on operating systems never stays static. Once secured, a system does not perpetually stay secure. Indeed, the longer one uses the system, the less secure it becomes. This can happen through operational or functional changes exposing the system to threats or through new exploits being discovered in packages and applications. Securing a system thus becomes an ongoing and living process.

*C. Current Solutions*

Established techniques to protect Internet-connected machines tend to rely either on filtering packets, or on application-level security. The first technique is implemented by firewalls, Internet-connected devices running software whose job is to filter or log unwanted network traffic. However, there are common attacks against which a firewall cannot protect. For example, firewalls do not protect against attempts to exploit bugs in application-level software. Such vulnerabilities occur because the Internet architecture assumes that services bound to a port should be accessible by any machine using the Internet protocols.

The second technique is to deploy high-strength application-level security mechanisms. However, authenticated services are built above the network layer and are often themselves subject to attack once discovered on a host. Reliance solely on application-level security exposes the problem of the computational expense of such security mechanisms. The high computational burden of commonly deployed cryptographic schemes leaves the server open to computational DoS attacks. Furthermore, complex schemes are error prone, as exemplified by the integer overflow bug in SSH which has been the target of many attacks [14].

*D. Attack Surface*

Recent trend [15, 16] towards mitigating security risks by minimizing the code you expose to the untrusted users has been gaining popularity. This philosophy is called Attack Surface Reduction (ASR). The attack surface of an app is the union of code, interfaces, services, protocols, and practices available to all users, with a strong focus on what is accessible to unauthenticated users. The core tenet of ASR is that all code has a nonzero likelihood of containing one or more vulnerabilities. Some vulnerabilities will result in system compromise. Therefore, the only way to ensure no system compromises is to reduce code usage to zero. ASR is a compromise between perfect safety and unmitigated risk that minimizes code exposed to untrusted users. Methods like Port Knocking and Single Packet Authentication (SPA) [17] are based on this concept. Port knocking's primary aim is to provide an extra layer of protection through the use of authentication with the added benefit of service cloaking. SPA retains all of the benefits of port knocking, but fixes the limitations of the latter by providing a stronger authentication mechanism.

Although both these methods allow administrators to keep the (potentially vulnerable) service hidden from the public, whilst making it available to authorized users, these share the problem that a stack access is required for the implementation to work. I.e. they rely on the services provided by the operating system, which may contain flaws.

### III ARCHITECTURE

*A. Overview*

The doctrine for setting up a secure network server, reflects on the core concepts discussed in the earlier section: service cloaking and authentication. The networked host should be minimalist in design i.e. only required services should be run and access to these services should only be provided to authenticated hosts. And this should be provided in a secure way without depending on other vulnerable software. One way of implementing this is to apply these security mechanisms in the network interface card itself thereby making it independent of OS and applications. This will not only secure the host from data link attacks but also will cover the applications running above it.

Thus, the requirements for protecting the host and the services running on it can be listed down as follows

- The services running on the machine should be hidden for normal hosts. This allows host server to attempt to be invisible to other Internet connected machines while still providing service to authorized parties.

- Only authorized source with valid credentials should be able to access the services. The authorization mechanism should be easy to validate, yet difficult to forge. For the service to remain hidden, this stage should elicit no response if the credentials are invalid.

- Full-strength application-specific security mechanisms can be used to augment the above mechanisms to provide true end-to-end encryption.

The benefits brought out from these schemes are:

- Reduced opportunity for attack attempts - since attacks are less likely to be attempted on machines which do not appear to be running any services;
- Reduced vulnerability to attacks- since an attacker must circumvent our scheme before proceeding with the attack;
- Reduced cost (in terms of processing time and resources) of rejecting attackers- since large fraction of packets are discarded at the interface itself, freeing the operating system of these procedures
- Immunity to data link attacks based on ARP poisoning- Since ARP is updated only after verification of the knock, any attempt to poison the ARP cache gets invalidated.

*B. Authentication*

Authentication is provided in the form of "Knock" sequences embedded in the packets. The packets contain the IP address and port number of the sender. The contents of the packet are encrypted using a shared key. This prevents any third party from extracting the IP address and port number of the sender, thereby disabling him from launching spoofing based attacks. The knock sequences are carried in ICMP packet which will allow external trusted hosts to use this mechanism.
The sequence of operations involved in establishing a secure connection is shown in Figure 1.
The sender first gets the MAC address of the server using normal ARP protocol. The reply in this case is not generated by the network stack of the server. An ARP Processor in the card performs this operation in stateless manner. This mechanism is explained in detail in Section 4. After receiving the MAC address the clients interface card generates a Knock packet. This is validated by the servers NIC and dynamically adds a filter table entry in the NIC, which will allow packet belonging to this particular transaction to reach the server

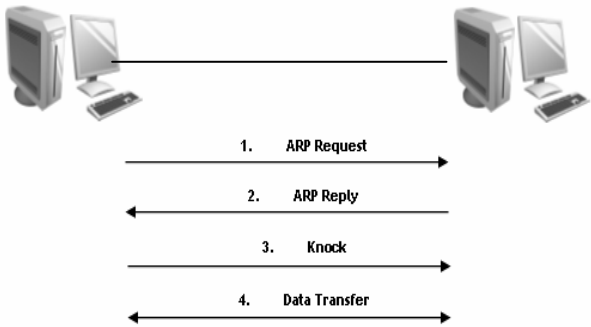

Figure 1. Packet Sequence

process. Since the IP address and port number that is opened is known only to the client and server (since the packet is encrypted), replay attacks are avoided. Once the filter rules are enabled, the client can carry on with the transaction. Detailed sequence of operation is depicted in Figure 2. The implementation details are provided in Section 4

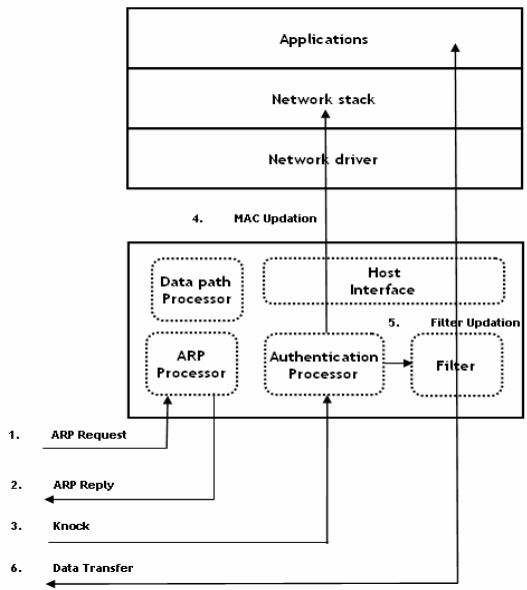

Figure 2. Sequence of operations

*C. Service cloaking*

Recent research [15, 16] has shown that rather than trying to minimize the 'exploit-disclosure-patch' cycle, a system's security can be 'virtually' increased via 'security through concealment' method. This method has the added benefit that the protected machines have a greater immunity against zero-day exploits. The crux of the idea is to minimize the Code Execution Paths (CEP), which is made visible by a system. Security measures like port knocking and Single Packet Authentication (SPA) are based on this model. Although the above methods are efficient enough in hiding the services, they too depend on the operating systems APIs. This makes them vulnerable to various exploits. For example both Port Knocking and Single Packet Authentication make use of NetFilter to deny the packets. Recent vulnerabilities found in the NetFilter implementation in Linux 2.6 kernels allow a remote host to hang a system running these versions of the software. By lowering the security mechanisms to a lower layer prevents these flaws. Figure 3 a,b,c compares the Code Execution Paths in various security implementations.

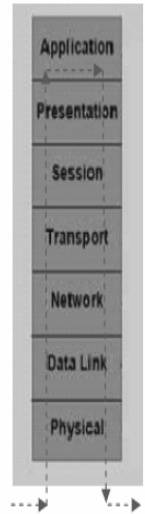 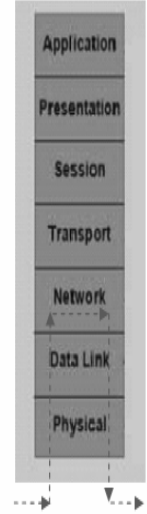 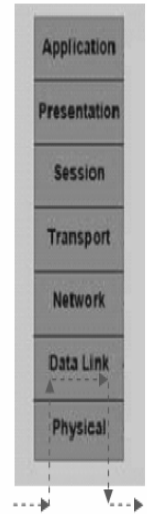

Figure 3a: CEP for application layer security

Figure 3b: CEP for network layer security

Figure 3c: CEP for data link layer security

The architecture also facilitates optional encryption of packet. The principal reason for not enforcing encryption is to allow network administrators and network devices (intrusion detection systems, anti-virus etc.) to look into the packets flowing in their network. This information is generally used to characterize the traffic to streamline the traffic flow across the network devices. Users, who still require a secure channel, can use any of the upper layer securities like IPsec or TLS.

**IV IMPLEMENATATION**

*A. Overview*

The card is designed in house as a PCI add-on board. It consists of an Altera CycloneII (EP2C672C7) FPGA (Figure 4). The Ethernet interfaces are provided through a pair of

LXT972 transceivers. The LXT972 is an IEEE compliant Fast Ethernet PHY Transceiver that directly supports both 100BASE-TX and 10BASE-T applications. It provides a Media Independent Interface (MII) for easy attachment to 10/100 Media Access Controllers (MACs). This interface is provided to the FPGA for Tx and Rx. Host connectivity is provided through a 32 bit 33 MHz PCI interface. The limit switches enable the card to work with 5v and 3.3v PCI connectors. The card also has external memory in the form of SDRAM (MT48LCM32B2). The SDRAM controller is a soft IP interfaces with the Avalon internal bus.

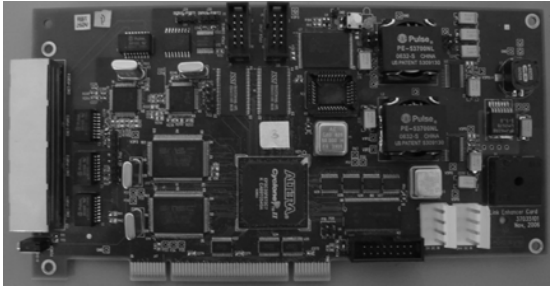

Figure 4. Card view

The Figure 5 shows the architecture of the firmware. A transceiver interface is used as a wrapper for the Ethernet Tx and Rx operation to the PHY device. The major modules include Data Path Processor, Authentication Processor, ARP Processor, Packet Filter and Host Interface. A set of asynchronous FIFOs are used to interface the PHY device. The FIFOs are 8bit wide and can store 2 maximum size Ethernet packets. These modules are further described below:

*B. Data Path Processor*

The Data Path Processor is implemented using the Nios II softcore processor. It is responsible for the overall control of the card. During transmission it generates the knock packet. It also processes the received knock packet. The MAC address of the validated knock along with the source IP obtained from the knock is provided to the host through the host interface for ARP cache updation. It also configures the filter table using the parameters obtained from the validated knock.

*C. Authentication Processor*

The Authentication Processor's job is to validate the knock packet that is obtained from the Rx FIFO. If the packet is validated its MAC address is passed on to the Data Path Processor to MAC updation. The IP source address and the port values obtained from the knock are also passed for Filter updation.

*D. ARP Processor*

ARP processor is responsible for providing replies to the ARP requests coming from other hosts. This process is offloaded from the network stack to protect the latter from various layer

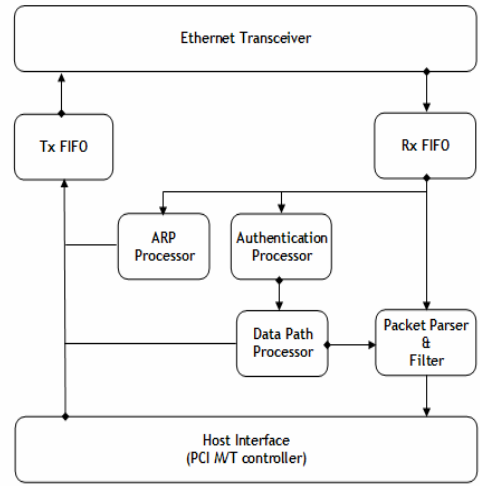

Figure 5 : Block diagram showing architecture of the firmware

2 attacks. The ARP processor uses the MAC address and IP address of the machine to create the ARP reply packet. The MAC address is hardwired and the IP address is obtained from the host PC during initialization.

*E. Packet Filter*

Packet Filter consists of a Packet Parser and a Filter Table. The default policy of the filter is to drop all packets coming to it. Packets that get a hit in the table are allowed to pass through. The table consists of the <IP source address, Source port number> vector.

*F. Host Interface*

Host interface is provided through a PCI Master/Target controller. The controller provides a 32 bit 33 MHz interface to the host machine. The internal address space is made available to the host through a set of Base Address Registers.

## V RELATED WORK

Sebatian in his thesis [17] on "Analysis of port knocking and Single Packet Authorisation" analyses the port knocking mechanism and single packet authorization. He assessed their suitability for firewall authentication mechanism for opening network ports. He also discusses the threats against port knocking. In [18], the author discusses an extension of port knocking mechanism called "Port hopping" based on the idea of Spread Spectrum frequency hopping.

Enhancing host based network security has been the subject of [19], [20] and [21], where the authors have used PCI based cards to improve the performance of existing software solutions.

## VI CONCLUSIONS AND FUTURE WORK

Despite many research and development efforts in the area of network security, it appears that the importance of data link layer security is still underestimated.

This paper presents a feasible solution to the problem of network security focusing on data link layer. The fundamental problems faced at the data link layer were discussed. It was shown that there is significant benefit by implementing security mechanisms at the data link layer. This considerably reduces the code execution path for most transactions thereby minimizing the attack vulnerability of the system. The said mechanisms were implemented in hardware and incorporated in a programmable network interface card.

However it is worth mentioning that there are a number of issues in the system that requires further considerations – like encryption and key management which are beyond the scope of this paper. These issues form the future work on the improvement and enhancement of the present implementation.

## REFERENCES

[1] TLS http://www.ietf.org/html.charters/tls-charter.html
[2] IPSec http://www.ietf.org/rfc/rfc2401.txt
[3] Lee Brotzman: Wrap a Security Blanket Around Your Computer, Linuxjournal article 1997-08-01
[4] Kevin J. Houle, George M. Weaver, CERT/CC, www.cert.org/archive/pdf/DoS_trends.pdf
[5] ARP Cache poisoning: http://www.grc.com/nat/arp.htm
[6] Code Red Worm, CERT Advisory CA-2001-19: http://www.cert.org/advisories/CA-2001-19.html
[7] Nimda Worm, CERT Advisory CA-2001-26: http://www.cert.org/advisories/CA-2001-26.html
[8] Common Vulnerabilities and Exposures: http://cve.mitre.org/
[9] Bugtraq : http://www.securityfocus.com/archive/1
[10] Anton Chuvakin: Linux Kernel Hardening, SecurityFocus article, http://www.securityfocus.com/infocus/1539
[11] IEEE 802.3 CSMA/CD (Ethernet), http://www.ieee802.org/3/
[12] Rinat Khoussainov, Ahmed Patel, LAN security: problems and solutions for Ethernet networks, http://www.smi.ucd.ie/~rinat/papers/Ethernet-security.pdf
[13] RFC 826, Ethernet Address Resolution Protocol, www.faqs.org/rfcs/rfc826.html
[14] OpenSSH Vulnerability, CERT Advisory CA-2002-18, http://www.cert.org/advisories/CA-2002-18.html
Denial of Service Attacks
http://www.cert.org/tech_tips/denial_of_service.html
[15] P. Manadhata and J. M. Wing, An Attack Surface Metric, Technical Report CMU-CS-05-155, July 2005.
[16] Michael Howard, Mitigate Security Risks by Minimizing the Code You Expose to Untrusted Users, MSDN Magazine, http://msdn.microsoft.com/msdnmag/issues/04/11/ AttackSurface/
[17] Sebatian Jeanquier," Analysis of port knocking and Single Packet Authorisation", MSc thesis report, Royal Holloway.
[18] Paul DiGioia, "Behind closed doors: An Evaluation of Port Knocking Authentication",Irvine, CA-92697-3425
[19] Arun Kumar S P et al. "NIC vs. Kernel based Filtering for Network Intrusion Detection Systems", In Proc. of First International Conference on Information Processing, 2007
[20] Haoyu Song, Todd Sproull, Mike Attig, John Lockwood: SNORT offloader: A Reconfigurable hardware NIDS filter,www.arl.wustl.edu/~hs1/publication/snort-offloader.pdf
[21] M. Otey et al.: NIC-based intrusion detection: A feasibility study, http://dmrl.cse.ohio-state.edu/papers/ICDM02-ws.pdf, November 27, 2002